\documentclass[11pt,a4paper]{article}                           

\usepackage{a4}          

\usepackage[UKenglish]{babel}
\usepackage{amsmath}
\usepackage{amssymb}
\usepackage[usenames]{color} 
\usepackage{multirow}
\usepackage{graphicx}
\usepackage{epstopdf}
\usepackage{array}
\usepackage{hhline}
\usepackage{cite}
\usepackage{microtype}
\usepackage[bf]{caption}
\usepackage{color}
\usepackage[usenames,dvipsnames]{xcolor}
\usepackage{subcaption}
\usepackage{pstool}

\usepackage{ifpdf}
\ifpdf
  \usepackage[pdftex,
    pdftitle={},
    pdfauthor={},
    pdfsubject={},
    bookmarksopen, bookmarksnumbered, bookmarksopenlevel=2]{hyperref}
\fi
\def\hybrid{\topmargin -20pt    \oddsidemargin 0pt
        \headheight 0pt \headsep 0pt
        \textwidth 6.25in       
        \textheight 9 in       
        \marginparwidth .875in
        \parskip 5pt plus 1pt 
          \jot = 1.5ex
   }
\hybrid
\numberwithin{equation}{section}
\numberwithin{table}{section}

\newcommand{\beq}{\begin{equation}}  \newcommand{\eeq}{\end{equation}}
\newcommand{\bal}{\begin{aligned}}   \newcommand{\eal}{\end{aligned}}
\newcommand{\bea}{\begin{eqnarray}}  \newcommand{\eea}{\end{eqnarray}}

\newcommand{\bmat}{\left(\begin{array}}
\newcommand{\emat}{\end{array}\right)}

\newcommand{\nn}{\nonumber}

\usepackage{cancel}

\newcommand{\be}{\begin{equation}}
\newcommand{\ee}{\end{equation}}

\newcommand{\half}{\frac{1}{2}}

\begin{document}

\baselineskip=14pt
\parskip 5pt plus 1pt

\vspace*{-1.5cm}
\begin{flushright}    
  {\small
  }
\end{flushright}

\vspace{2cm}
\begin{center}        
  {\LARGE The Weak Gravity Conjecture and Scalar Fields}
\end{center}

\vspace{0.5cm}
\begin{center}        
{\large Eran Palti}
\end{center}

\vspace{0.15cm}
\begin{center}        
  \emph{Institut f\"ur Theoretische Physik, Ruprecht-Karls-Universit\"at, 
             Heidelberg, Germany}
             \\[0.15cm]
 
\end{center}

\vspace{2cm}


\begin{abstract}
We propose a generalisation of the Weak Gravity Conjecture in the presence of scalar fields. The proposal is guided by properties of extremal black holes in ${\cal N}=2$ supergravity, but can be understood more generally in terms of forbidding towers of stable gravitationally bound states. It amounts to the statement that there must exist a particle on which the gauge force acts more strongly than gravity and the scalar forces combined. We also propose that the scalar force itself should act on this particle stronger than gravity. This implies that generically the mass of this particle decreases exponentially as a function of the scalar field expectation value for super-Planckian variations, which is behaviour predicted by the Refined Swampland Conjecture. In the context of ${\cal N}=2$ supergravity the Weak Gravity Conjecture bound can be tied to bounds on scalar field distances in field space. Guided by this, we present a general proof that for any linear combination of moduli in any Calabi-Yau compactification of string theory the proper field distance grows at best logarithmically with the moduli values for super-Planckian distances.
 \end{abstract}

\thispagestyle{empty}
\clearpage

\setcounter{page}{1}


\newpage

\tableofcontents

\section{Introduction}

It is expected that not all consistent Quantum Field Theories (QFTs) can arise as effective theories of ultraviolet physics which includes quantum gravity. Understanding the precise nature of the constraints on QFTs is both theoretically interesting and can have implications for phenomenological model building.

One such quantitative constraint is the Weak Gravity Conjecture (WGC) \cite{ArkaniHamed:2006dz}. It states that in a theory with a $U(1)$ gauge symmetry, with gauge coupling $g$, there must exist a state of charge $q$ and mass $m$ which satisfies the inequality
\be
g q M_p \ge m \;. \label{WGC}  
\ee
For this particular state gravity acts equally or more weakly than the $U(1)$ force. The WGC, and various modifications of it, have been studied intensively, see \cite{Cheung:2014vva,Rudelius:2015xta,Brown:2015iha,Montero:2015ofa,Hebecker:2015rya,Brown:2015lia,Heidenreich:2015wga,Palti:2015xra,Heidenreich:2015nta,Kooner:2015rza,Ibanez:2015fcv,Hebecker:2015zss,Hebecker:2015tzo,Conlon:2016aea,Heidenreich:2016aqi,Montero:2016tif,Hebecker:2016dsw,Saraswat:2016eaz,Herraez:2016dxn,Cottrell:2016bty,Hebecker:2017uix,Hebecker:2017wsu,Dolan:2017vmn} for the most recent results. The bound (\ref{WGC}) was motivated in \cite{ArkaniHamed:2006dz} by the requirement that extremal black holes should be able to decay. It was also motivated through the statement that the particle with the largest charge to mass ratio should not form gravitationally bound states. Such states would be stable against decay by charge and energy conservation. It was argued in \cite{ArkaniHamed:2006dz} (see also \cite{Cottrell:2016bty}) that such a tower of stable states would be problematic for quantum gravity. The absence of such stable gravitationally bound states will play an important role in our analysis.

While the decay of extremal black holes and no stable gravitationally bound states are general principles, the formulation (\ref{WGC}) is an application of them to simple Reissner-Nordstrom (RN) black holes where only gravity and gauge fields play a role. In this paper we will be interested in how the WGC is modified when scalar fields are present. We will consider the most general (two-derivative) action of massless bosonic fields
\be
{\cal L} \left(-g\right)^{-\frac12} = \frac{R}{2} - g_{ij}\left(t\right)\partial_{\mu} t^i \partial^{\mu} t^{j} + {\cal I}_{IJ}\left(t\right) {\cal F}_{\mu\nu}^{I} {\cal F}^{J,\mu\nu} +  {\cal R}_{IJ}\left(t\right) {\cal F}_{\mu\nu}^{I} \left(\star {\cal F}\right)^{J,\mu\nu} \;,    \label{genaction}
\ee
with scalar fields $t^i$ and gauge fields $A^I$, and propose the general formulation of the Weak Gravity Conjecture for this. We first formulate it in an ${\cal N}=2$ supergravity context through properties of extremal black holes, but propose that it generalises and can be understood as the general statement forbidding stable gravitationally bound states. In short, it amounts to the statement that there must exist a state on which the gauge force should be stronger than gravity and the forces mediated by the scalar fields combined. 

Another conjecture, termed the Refined Swampland Conjecture (RSC) \cite{Baume:2016psm,Klaewer:2016kiy}, states that once a scalar field varies over a super-Planckian distance $\Delta \phi \ge M_p$ there is an infinite tower of states, with mass scale $m$, whose mass decreases exponentially fast as a function of the scalar field variation
\be
m\left( \phi + \Delta \phi \right) \leq  m\left( \phi \right) e^{-\alpha \frac{\Delta \phi}{M_p}} \;, \label{RSC} 
\ee
where $\alpha$ is a constant which is determined by the choice of direction of the variation in field space. The conjecture is based on an earlier weaker statement in \cite{Ooguri:2006in} about infinite distances in moduli space which we term the Swampland Conjecture, and is supported primarily by evidence from string theory \cite{Ooguri:2006in,Baume:2016psm,Valenzuela:2016yny,Blumenhagen:2017cxt}.\footnote{There are a number of papers studying closely related questions regarding super-Planckian variations in string theory, see \cite{Hebecker:2015tzo,Landete:2016cix,Bielleman:2016olv,Landete:2017amp} for the most recent work.} In \cite{Klaewer:2016kiy} a general argument for the RSC applied to fields which appear in a gauge coupling of a $U(1)$ was presented based on black hole physics. Nonetheless, the RSC has not yet been understood in terms of a general principle in the same sense as the WGC.

In \cite{Baume:2016psm} it was pointed out that there exists a simple relation between field variations and the RSC on one hand and the WGC on the other which arises in the presence of supersymmetry. Its simplest formulation utilises the WGC applied to axions, rather than gauge fields, combined with ${\cal N}=1$ supersymmetry. The WGC as applied to axions states that $f S \leq M_p$, where $f$ is the axion decay constant and $S$ is the action of the instanton coupling to the axion. In the presence of supersymmetry both of these quantities are related to properties of the axion scalar superpartner, the saxion denoted $t$, so that the inequality can be written as $\sqrt{g_{tt}} t \leq M_p$. Here $g_{tt}$ is the metric on the single saxion field space. This implies that for $t \geq M_p$ the proper field distance, as measured by the canonically normalised field, grows at best logarithmically with $t$. The logarithmic growth is further tied to the exponential behaviour of the RSC (\ref{RSC}) if the mass of the tower of states behaved as a power law in $t$. 

In this paper we generalise this simple argument and show that indeed the WCG and the RSC are, at least for certain scalar fields, in some sense superpartners. We consider the framework of ${\cal N}=2$ supergravity and show that the same ${\cal N}=2$ identity which leads to the general expression of the WGC also leads to a bound on the growth of proper distances in field space. As an application, this will allow us to prove that any linear combination of geometric moduli fields in Calabi-Yau or Calabi-Yau orientifold compactifications of string theory has a proper field distance which grows at best logarithmically at super-Planckian values. We also further identify a tower of states which becomes exponentially light as a result of this logarithmic behaviour, providing new evidence for the RSC. 

Within the ${\cal N}=2$ supergravity framework we also show that there must exist states on which the scalar field forces themselves act stronger than gravity. We consider the possibility that this is a general statement relating to bound states, but find that the arguments for this require further work to establish clearly. We show that requiring the scalar force to act stronger than gravity suggests that the WGC states depend exponentially on the scalar fields for super-Planckian scalar field variations, which is the statement of the RSC. Applying it to axionic fields leads to evidence that the WGC state is part of an infinite tower. 

\section{The Weak Gravity Conjecture with Scalar Fields}
\label{sec:n2ebh}

In this section we propose a generalisation of the WGC to the case when scalar fields are present. Since the WGC is tied to extremal black holes, to first build up the intuition we need to consider a large class of black holes in the presence of scalar fields. We do this in the context of ${\cal N}=2$ supergravity where the framework of extremal black holes is well understood. We then propose a generalisation of the WGC, independent of supersymmetry, inspired by the black hole physics.

\subsection{${\cal N}=2$ Extremal Black Holes and the Weak Gravity Conjecture}

Extremal black hole solutions of ${\cal N}=2$ supergravity have been extensively studied. See \cite{DallAgata:2011zkh} for a review. The action takes the form
\be
\left(-g\right)^{-\frac12} {\cal L}  = \frac{R}{2} - g_{ij}\partial_{\mu} z^i \partial^{\mu} \overline{z}^{j} + {\cal I}_{IJ} {\cal F}_{\mu\nu}^{I} {\cal F}^{J,\mu\nu} +  {\cal R}_{IJ} {\cal F}_{\mu\nu}^{I} \left(\star {\cal F}\right)^{J,\mu\nu} \;.     \label{N2action}
\ee
Here $R$ denotes the Ricci scalar and we set $M_p=1$. The ${\cal F}^I$ are the electric field strengths of $U(1)$ fields and the magnetic field strengths ${\cal G}_I$ are defined as
\be
{\cal G}_I = -\frac{\delta {\cal L}}{\delta {\cal F}^I} = {\cal R}_{IJ} {\cal F}^J  -  {\cal I}_{IJ} \star {\cal F}^J \;,
\ee
where $\star$ denotes the Hodge star. The $z^i$ are complex scalar fields, with field space metric $g_{ij}$, which have components 
\be
z^i = b^i + i t^i \;. \label{zt}
\ee
The indices are ranged such that $I=0,...,n_V$, and $i=1,...,n_V$, where $n_V$ is the number of vector multiplets. The geometric structure on the field space is determined through the periods $\left\{X^I,F_I\right\}$ which are holomorphic functions of the scalar fields $z^i$. These are related through a symplectic matrix $F_I = {\cal N}_{IJ} X^J$, which by supersymmetry determines ${\cal I}_{IJ} =\mathrm{Im} \left[{\cal N}_{IJ} \right]$ and ${\cal R}_{IJ} =\mathrm{Re} \left[{\cal N}_{IJ} \right]$. The Kahler potential for the scalar field-space metric takes the form 
\be
K = -\ln i\left( \overline{X}^I F_I - X^I \overline{F}_I \right) \;. \label{N2KahlerPot}
\ee
In certain cases it is possible to go to special coordinates $X^I=\left\{1,z^i\right\}$. In those cases the periods are also determined in terms of a prepotential $F$ through $F_I=\partial_{X^I} F$, and the general expression for the symplectic matrix takes the form
\be
{\cal N}_{IJ} = \overline{F}_{IJ} + 2i \frac{\mathrm{Im} F_{IK} \mathrm{Im} F_{JL} X^K X^L }{\mathrm{Im} F_{MN} X^MX^N} \;,
\ee
where $F_{IJ} = \partial_I F_J$.

It is useful to introduce 
\be
{\cal M} \equiv \left( \begin{array}{cc} {\cal I} + {\cal R} {\cal I}^{-1} {\cal R}& - {\cal R} {\cal I}^{-1} \\ -{\cal I}^{-1} {\cal R} & {\cal I}^{-1} \end{array} \right) \;, \;\; {\cal Q} \equiv \left( \begin{array}{c} p^I \\ q_I \end{array} \right) \;. \label{Mdef}
\ee
and the notation
\be
{\cal Q}^2 \equiv -\frac12 {\cal Q}^T {\cal M} {\cal Q} \;, \;\; {\cal Q}{\cal Q}' \equiv -\frac12 {\cal Q}^T {\cal M} {\cal Q'}  \;. \label{qdefq}
\ee
There is an identity which will play a central role in our analysis \cite{Ceresole:1995ca}\footnote{From the perspective of string theory there is an intuitive way to understand this structure within the context of flux compactifications of type II string theories on Calabi-Yau manifolds. For example, in type IIB both the $U(1)$ field strengths and three-form NS fluxes arise from reduction on three-cycles. The right hand side of (\ref{si}) can then be understood as the ${\cal N}=1$ formula for the flux induced scalar potential in terms of the superpotential. The important point is that the Kahler moduli and dilaton do not appear in the superpotential and therefore obey a no-scale type relation such that their F-terms are equal to $4\left|Z\right|^2$.}
\be
{\cal Q}^2 = \left|Z\right|^2 + g^{ij} D_i Z \overline{D}_j \overline{Z} \;. \label{si}
\ee
Here $q_I$ and $p^I$ are arbitrary constants, $Z$ is the central charge 
\be
Z = e^{\frac{K}{2}} \left(q_I X^I  - p^I F_I \right) \;.
\ee
The covariant derivative acts as
\bea
D_i \psi^j &=& \partial_{z_i} \psi^j  + \Gamma^{j}_{ik} \psi^k  + \frac{p}{2}\left(\partial_{z_i} K \right) \psi^j \;,
\eea
on an object $\psi^j$ with Kahler weight $p$ ($Z$ has weight 1).

We are interested in black hole solutions to the action (\ref{N2action}). The black hole electric and magnetic charges $\left(Q^I,P^I\right)$, as defined through the integration over a sphere at infinity,
\be
\frac{1}{4\pi} \int_{S_{\infty}} F^I =  P^I \;,\;\; \frac{1}{4\pi} \int_{S_{\infty}} \star F^I = Q^I \;,
\ee
 are related to quantised symplectic charges $\left(q_I,p^I\right)$ through
 \be
\left( \begin{array}{c} P^I \\ Q^I \end{array} \right)  =  \left( \begin{array}{c} p^I \\ \left( {\cal I}^{-1} \cdot {\cal R} \cdot p \right)^I - \left({\cal I}^{-1}\cdot q \right)^I \end{array} \right) \;.
\ee
The ${\cal N}=2$ supersymmetric extremal black hole solutions with charges $\left(q_I,p^I\right)$ have an ADM mass given by the central charge 
\be
M_{\mathrm{ADM}} = \left|Z\right|_{\infty} \label{admmass}  \;.
\ee
The subscript $\infty$ denotes the evaluation of the fields at their values at spatial infinity. We will often drop this index leaving this implicit when the setting is sufficiently clear.

The WGC as stated in (\ref{WGC}) arises from RN black holes which, in the electric case, satisfy $g^2 q^2 = M^2_{\mathrm{ADM}}$. We would like to write a generalisation of this for general ${\cal N}=2$ extremal black holes. To do this we must construct a field reparametrisation invariant quantity. Using (\ref{admmass}) and (\ref{si}) leads to the natural generalisation 
\be
{\cal Q}^2  = M_{\mathrm{ADM}}^2 + g^{ij} D_i Z \overline{D}_j \overline{Z} \;. \label{WGCadmmass}
\ee
The attractor mechanism states that the values of the scalar fields on the horizon of an extremal black hole are fixed in terms of the black hole charges as $D_i Z = 0$. There are therefore two types of extremal black holes. The first are where the scalar field values at infinity differ from those on the black hole horizon so that there is a scalar field spatial gradient. The second type, denoted double extremal black holes \cite{Kallosh:1996tf}, are where the values at infinity are equal to those on the horizon and the scalar fields have a constant spatial profile. Such double extremal black holes therefore maximise the ADM mass of the black hole relative to its charge at infinity. 

It is informative to rewrite (\ref{WGCadmmass}) as
\be
{\cal Q}^2 = M_{\mathrm{ADM}}^2 + 4 g^{ij} \partial_i M_{\mathrm{ADM}} \overline{\partial}_j M_{\mathrm{ADM}} \;. \label{gWGCmadm}
\ee
Here, since we have derivatives acting on $M_{\mathrm{ADM}}$, we should think of it as the mass as a function of the moduli $z^i$, rather than just in the vacuum. 

We would now like to follow the logic that particles should exist for the black hole to be able to decay. This can be applied in the case of ${\cal N}=2$ supergravity, but the presence of extended supersymmetry means that the black hole decay properties are highly restricted. The supersymmetric black holes are BPS states and this means that they are at best marginally stable, and sometimes only over special loci in field space termed curves of marginal stability. When they do decay, it can only be to other BPS states. 

The last term of (\ref{gWGCmadm}) is positive definite. Therefore, for these extremal black holes to be able to decay we can impose that there must exist a particle with mass $m$ such that ${\cal Q}^2 \geq m^2$. However, if the last term in (\ref{gWGCmadm}) is non-vanishing we find a stronger constraint that the particle must be strictly super-extremal. Indeed, since the particle must itself be a BPS state, its mass is given by the central charge and therefore it satisfies 
\be
{\cal Q}^2 \geq m^2 + 4 g^{ij} \partial_i m \overline{\partial}_j m \;.  \label{gWGCN2}
\ee
In fact, since it is BPS, the inequality in (\ref{gWGCN2}) is saturated. However, in the next section we will utilise it as an inequality for more general, and possibly non-supersymmetric, cases. 

\subsection{Generalisation to non-Supersymmetric Theories}

It is interesting to consider if an analogous bound to (\ref{gWGCmadm}) holds for black holes which are only half-BPS or even not supersymmetric at all. It is possible to write for any extremal black hole a black hole scalar potential which is the gauge kinetic terms as a function of the scalar fields. This is precisely ${\cal Q}^2$ appearing in (\ref{WGCadmmass}). If this scalar potential can be written as
\be
V_{\mathrm{BH}} = {\cal Q}^2 = {\cal W}^2 + 4g^{ij} \partial_i {\cal W} \overline{\partial}_j {\cal W} \;, \label{bhfs}
\ee
where ${\cal W}$ is a real function, termed the `fake superpotential', then the ADM mass is given by $M_{\mathrm{ADM}} = \left.{\cal W}\right|_{\infty}$ and on the horizon the fields solve $\partial_i {\cal W}=0$ \cite{Denef:2000nb,Ceresole:2007wx}. This is suggestive that the bound (\ref{gWGCmadm}) is tied to extremality rather than supersymmetry. 

Even if (\ref{gWGCmadm}) holds generally, it does not imply (\ref{gWGCN2}), at least not utilising only charge and energy conservation. The requirement for decay can be stated as the existence of a particle with a larger charge to mass ratio than the black hole. This can be written as\footnote{At least for a single $U(1)$ and decays to electric particles (\ref{nonsusydecay}) follows from the usual arguments of charge and mass conservation. By completing this to a field reparametrisation invariant quantity it is natural to impose (\ref{nonsusydecay}) more generally, however we did not derive this and leave such a result for further work. It is useful to note through that if the charge vectors of the decay particles all have positive or vanishing $Q \tilde{Q}$, as in (\ref{qdefq}), for any pair, then (\ref{nonsusydecay}) again follows simply.}
\be
\left( \frac{{\cal Q}^2_{\mathrm{BH}}}{M^2_{ADM}} \right) = 1 + 4 g^{ij} \partial_i \ln M_{ADM} \bar{\partial}_j \ln M_{ADM} \leq \left( \frac{{\cal Q}^2_{\mathrm{Particle}}}{m^2} \right) \;. \label{nonsusydecay}
\ee
Here ${\cal Q}^2_{\mathrm{BH}}$ and ${\cal Q}^2_{\mathrm{Particle}}$ are the relevant expressions for the black hole and the particle, with respective masses $M_{ADM}$ and $m$. While the particle is required to be super extremal, the particular expression (\ref{gWGCN2}) would require a replacement of $M_{ADM}$ with $m$ in (\ref{nonsusydecay}). It is natural to expect that the charge to mass relation for the particle should indeed only involve $m$ and not $M_{ADM}$, and that it should form a field reparametrisation invariant quantity, which motivates a form (\ref{gWGCN2}). Further, as we change the values of the scalar fields at infinity $M_{ADM}$ also changes, and in order for the particle to maintain a decay channel its mass needs to also change appropriately. This relation between the functional field dependence of the black hole and the particle can motivate a relation between (\ref{nonsusydecay}) and (\ref{gWGCN2}). However, in the absence of ${\cal N}=2$ supersymmetry, it is unclear how to make a sharp general argument for (\ref{gWGCN2}), rather than (\ref{nonsusydecay}), using black hole decay. 

To further motivate a general statement we can consider the physics captured by (\ref{gWGCmadm}) and (\ref{gWGCN2}). The inequality (\ref{gWGCN2}) can be phrased as the statement that the $U(1)$ force between two WGC states acts at least as strongly as the gravitational and scalar forces combined. This can be seen by noting that the last term in (\ref{gWGCN2}) is the force due to the exchange of the scalar fields $z^i$ induced by the cubic coupling of these fields to two WGC states from the mass term.\footnote{Note that in going from the cubic coupling in the Lagrangian to the classical force coupling there is a factor of two times the mass of the external states due to the change from relativistic to non-relativistic wavefunction normalisation.} The scalar contribution can be simply calculated in the usual way as the potential induced through the exchange diagram of the $z^i$. The gauge force contribution ${\cal Q}^2$ is substantially more complicated than the simple case $g q$ in (\ref{WGC}), but this is due to it being a general expression for dyonic objects and in the background of a non-vanishing $\theta$-angle matrix ${\cal R}_{IJ}$. For BPS states the inequality becomes an equality which is the expression of the no-force condition. 

In the context of the WGC this can be understood as a generalisation of the statement that gravity is the weakest force. It amounts to forbidding gravitationally bound states of the WGC states. Such states will be stable if the WGC states are those with the smallest mass to charge ratio. The existence of such a tower of stable states was argued to be problematic in \cite{ArkaniHamed:2006dz} and \cite{Cottrell:2016bty}. The requirement of the absence of such bound states will play a central role in our analysis, however, we will not focus on justifying this requirement as a property of quantum gravity, but will assume it and study its consequences. The scalar field contribution is positive definite on the side of gravity. This is because the scalar force between equal charged particles is attractive. The positivity of the sum over all the scalar forces is then ensured by the fact that the field space metric $g_{ij}$ must be positive definite. Therefore, we see that the absence of a bound state means that the gauge field repulsion must overcome both the gravitational and scalar field attraction. 

Given this general understanding we can formulate the conjecture more generally. Consider a theory with gauge kinetic matrices ${\cal I}_{IJ}$ and ${\cal R}_{IJ}$, and massless real scalar fields $t^i$ with field space metric $g_{ij}$ as in (\ref{genaction}). There need not be any relation, including the index ranges, between the gauge kinetic matrices and the field space metric. The conjecture is then that there must exist a particle, with mass $m$, satisfying a bound
\be
{\cal Q}^2 \geq m^2 + g^{ij} \mu_i \mu_j \;. \label{gnsWGC}
\ee
Here $\mu_i$ are the (non-relativisitic) couplings of the WGC state to one $t^i$. If we write the mass as a function of the fields $t^i$ then
\be
\mu_i = \partial_{t^i} m \;.
\ee
The inequality (\ref{gnsWGC}) ensures that there are no gravitationally bound states of the particle. The existence of a particle satisfying (\ref{gnsWGC}) is motivated by considering the particle with the smallest mass to charge ratio in which case a gravitationally bound state is stable. There are stronger versions of the WGC which state that the particle may satisfy other criteria. For example, that it should hold for the particle of minimal charge \cite{ArkaniHamed:2006dz}, and the generalisation of this that it should hold for an infinite tower of particles which is the Lattice WGC \cite{Heidenreich:2015nta,Heidenreich:2016aqi}. In string theory it appears that both of these are true, and it is certainly conceivable that (\ref{gnsWGC}) should indeed hold for these stronger constraints on the properties of the particles. 

A natural question which arises is whether we should take (\ref{gnsWGC}) to hold over all of the scalar field space or only certain loci? We can utilise the ${\cal N}=2$ setting to gain some intuition. A BPS black hole can only decay over loci of marginal stability. If the charges of the black hole are relatively prime then this is a strict sub-locus of the full field space. This suggests that perhaps we should impose (\ref{gnsWGC}) only over sub-loci of the field space. However, the BPS nature of the states that the black hole decays to ensures that they satisfy an equality version of (\ref{gnsWGC}) over all field space. Therefore, it natural to expect that (\ref{gnsWGC}) should be taken to hold over all of field space, but that this does not necessarily mean that the extremal black hole can decay to the particle at all points in field space. A more refined conclusion can be reached by thinking about the existence of stable gravitationally bound states. The particle with the largest charge to mass ratio should not form a gravitationally bound state since it would be stable. As we move around field space it could happen that a different particle becomes the one with the largest charge to mass ratio. Then now we can allow for gravitationally bound states of the original particle since they could decay to the new particle. Therefore, the natural conclusion is to impose that at any point in field space there is one state which satisfies (\ref{gnsWGC}) but that this may not be the same state over all of field space. In the presence of multiple gauge fields it is natural to expect that the transition between WGC states should correspond to loci of marginal stability. In our analysis we will assume that indeed (\ref{gnsWGC}) should hold for at least one state at any point in field space, but note that in terms of requiring black hole decay there is a possibility that it should only hold over sub-loci of field space. It is also worth noting that in any single direction in charge space the loci of marginal stability are just the whole field space, since the charges are not relatively prime, and so these subtleties do not play a role.

The minimal requirement, in the presence of $N$ $U(1)$s, is that the relation (\ref{gnsWGC}) should apply to $N$ particles with charge vectors spanning an $N$-dimensional space. In the absence of scalar fields, the charge (over mass) vectors needed to be such that their convex hull includes the unit circle \cite{Cheung:2014vva}. The unit circle is the configuration space of RN black hole. However, for more general black holes this space is very different. It is unclear how to formulate generally a natural stronger statement than the minimal one. Certainly one possibility is to require that a WGC state exists for every charge choice, which is the Lattice WGC \cite{Heidenreich:2015nta,Heidenreich:2016aqi}. 

Note that the bound (\ref{gnsWGC}) implies that as the strength of the gauge coupling goes to zero not only does the mass of the WGC state vanish but also all its interactions with any scalar field. This appears to be a strong statement in the sense that the weak coupling limit $g \rightarrow 0$ is leading to quite drastic behaviour. This is consistent with the picture of the magnetic WGC of $g$ forming a cutoff mass scale of the theory. 

If we consider the state with the largest charge to mass ratio to be light, so send $m \rightarrow 0$, we recover a constraint on the charges and couplings of light states. This imposes non-trivial constraints on field theories at energy scales far below the Planck scale. It is also interesting to note that if there is a cancellation between the gauge and scalar forces to a large degree, then it forces the mass scale of the state to be very light. 

The analysis so far has applied to massless gauge fields and massless scalar fields. In light of the possible infrared implications of  (\ref{gnsWGC}), it is interesting to consider how it would be modified for massive force mediators. Consider how a theory consistent with the WGC should behave. We start with massless gauge and scalar fields. We then deduce a constraint as in (\ref{gnsWGC}). Now we give the gauge or scalar fields a small mass. It seems that as long as this mass if sufficiently smaller than the mass of the WGC state it should not modify the mass of the WGC state. Similarly, it seems unlikely that it would modify the coupling to scalar fields. Therefore, if the mass of the gauge or scalar fields is below the mass scale of the WGC state we may expect that (\ref{gnsWGC}) should hold.

However, it is unclear if an analysis of bound states supports this expectation. The classical long range force analysis relies on taking the mass of the force mediators much smaller than that of the WGC states. We can consider first the possibility that the repulsive force carrier has a mass. At sufficiently large distances gravity will always beat a massive force carrier. Therefore, if the repulsive force mass, in this case the gauge field mass, is non-vanishing we will always have bound states.\footnote{In section \ref{sec:refgrw} we will also consider repulsive scalar forces, and the analysis will apply equally then.} It is then natural to wonder what the role of (\ref{gnsWGC}) would be. In this case it would ensure that the length scale of these states will behave like the inverse mass of the carrier. Also this length scale will increase as we build up charge. This is in contrast to the case when (\ref{gnsWGC}) is violated in which case their length scale can be arbitrarily small. So if we give a small mass to the carrier then as long as we satisfy (\ref{gnsWGC}) we do have bound states but they are over very large distances which seems less problematic from a quantum gravity perspective. Having stated this, by taking a sufficiently large number of the WGC states it appears to be possible to form a black hole with a radius much larger than the scale at which the repulsive force can act. It is unclear to us what the implications of this are. 

If the attractive scalar force gains a mass then we could imagine violating  (\ref{gnsWGC}) and still not forming bound states at scales much larger than the inverse mass. However, we could form bound states at arbitrarily small distance scales. Such states would only be classically bound through a barrier rather than by charge and mass conservation, so it is not clear if they are problematic. 

It is therefore difficult to reach a conclusive statement on if, and how, (\ref{gnsWGC}) should be modified when the gauge or scalar fields have a mass. It would be very interesting to try and understand this better as it could have important implications for the infrared consequences of (\ref{gnsWGC}).

Of course, a natural question which arises is whether gravity acts weaker than the scalar field forces themselves. We explore this in section \ref{sec:refgrw}.

It is interesting to invert the reasoning and utilise the logic that (\ref{gnsWGC}) captures the fact that there should be no stable bound states to impose a constraint on the structure of extremal black holes. This appears to suggest that an expression involving a fake superpotential, as in (\ref{bhfs}), should be general for extremal black holes.  

\section{Distances in Moduli Space}
\label{sec:distmod}

Having studied how the WGC is formulated in the presence of scalar fields, we now turn to the seemingly unrelated topic of distances in field space. This is the topic of the Refined Swampland Conjecture as stated in (\ref{RSC}). There is a closely related conjecture made in \cite{Baume:2016psm} that field distances grow logarithmically for super-Planckian distances. Of course, logarithmically is a meaningless statement without specifying what it is logarithmic in. In \cite{Baume:2016psm} the fields were closed string axions, and the proper field distance behaved logarithmically in the field obtained by reducing the higher dimensional Ramond-Ramond and Neveu-Schwarz fields on a cycle. In this section we will primarily consider moduli fields of Calabi-Yau compactifications, and aim to show that proper field distances grow logarithmically in the fields obtained by reducing the Kahler $J$ and holomorphic three-form $\Omega$ on cycles, or in other words, in the volume of the cycles.

It is well known that in simple setups, such as torodial compactifications, it is indeed the case that field distances grow logarithmically in the moduli. However, there are no general results for more complicated constructions such as Calabi-Yau compactifications. Given the complexity of a typical Calabi-Yau moduli space, if a general result exists then it is likely that there is an underlying general reason for this. As discussed in the introduction, in \cite{Baume:2016psm} it was pointed out that a simple relation between field distances and the WGC arises in the presence of supersymmetry. This relation will form a guiding principle for the analysis in this section. Logarithmic growth of field distances is not sufficient to prove the RSC. One still needs to show that there is an infinite tower of states whose mass depends, as a power law, on the moduli. This will be the topic of section \ref{sec:masssta}.

\subsection{Logarithmic Growth of Field Distances}
\label{sec:loggro}

Consider a field space spanned by real fields $t^i$. We would like to analyse a proper field distance along a direction in field space. Consider this direction to be of the form 
\be
\rho = \sum_i h_i t^i \;.
\ee
Here the $h_i$ are arbitrary constants but we can, with full generality for the purpose of analysing distances in field space, normalise them such that 
\be
\mathrm{Max}\left\{ \left|h_i\right| \right\}=1 \;. \label{hnorm}
\ee
The proper distance along this direction can be written as
\be
\Delta \phi = \int_{\rho_i}^{\rho_f} \left(h_i g^{ij} h_j \right)^{-\half}  d \rho \;, \label{defphprsim}
\ee
where $g_{ij}$ is the metric on the field space and the field $\rho$ varies from its initial values $\rho_i$ to its final one $\rho_f$. We present a derivation of (\ref{defphprsim}) in appendix \ref{sec:derv}.

Evaluating (\ref{defphprsim}) is a difficult task. Even for a small number of fields inverting the field space metric is a non-linear problem which quickly becomes intractable. If we consider that a typical Calabi-Yau moduli space can contain hundreds such fields it is clear that forcing an explicit calculation of (\ref{defphprsim}) in such cases is not feasible. 

In the previous section we showed that the generalisation of the WGC to ${\cal N}=2$ systems is based on the identity (\ref{si}). With the relation between the WGC and field distances discussed at the start of this section in mind, we can look to (\ref{si}) for a possible result on field distances. We would primarily like to obtain a result for moduli fields in Calabi-Yau compactifications of string theory. In this case both the Kahler and complex structure moduli span a moduli space supporting so-called special geometry. This mean that they can be described in an ${\cal N}=2$ framework with a cubic prepotential. In the so-called large volume regime, which is $t^i \gg 1$, the prepotential takes the form
\be
F = -\frac{1}{6} K_{ijk}\frac{X^i X^j X^k}{X^0} \;. \label{cupre}
\ee
Given this prepotential we have (see for example \cite{Grimm:2005fa})
\bea
K &=& - \ln \frac{4\kappa}{3}\;, \nn \\
g_{ij} &=&  -\frac{3}{2\kappa} \left(\kappa_{ij} - \frac{3 \kappa_i \kappa_j}{2 \kappa} \right) \;, \;\; 
g^{ij} =  -\frac{2\kappa}{3}\left(\kappa^{ij} - \frac{3 t^i t^j}{\kappa} \right) \;, \\
{\cal I}_{IJ} &=& -\frac{ \kappa}{6} \left( \begin{array}{cc} 1 + 4g_{ij}b^ib^j  & - 4 g_{ij}b^j \\ - 4 g_{ij}b^j & 4 g_{ij} \end{array}\right) \;,\;\; 
{\cal I}^{IJ} = -\frac{6}{\kappa} \left( \begin{array}{cc} 1 & b^i \\ b^i & \frac14 g^{ij} + b^ib^j \end{array}\right) \;.
\eea
Here we define
\be
\kappa = K_{ijk}t^i t^j t^k \;,\;\; \kappa_i = K_{ijk}t^j t^k \;,\;\; \kappa_{ij} = K_{ijk}t^k \;.
\ee
Supersymmetry relates the inverse gauge kinetic metric ${\cal I}^{IJ}$ to the inverse field space metric $g^{ij}$. Therefore, by choosing the charges $q_I = \left(0,h_i\right)$ and $p^I=0$ in (\ref{si}), and setting the axions to zero $b^i=0$, we obtain the relation\footnote{We are free to set $b^i=0$ since the metric $g_{ij}$ does not depend on the $b^i$. }
\be
h_i g^{ij} h_j =  \frac{4\kappa}{3} \left( \left|Z\right|^2 + g^{ij} D_i Z \overline{D}_j \overline{Z}\right)  = \rho^2 +   \frac{4\kappa}{3}  g^{ij} D_i Z \overline{D}_j \overline{Z}  \;. \label{ins}
\ee
The crucial property of (\ref{ins}) is that the last term is positive definite, this is all we need to use and so let us write (\ref{ins}) as
\be
h_i g^{ij} h_j  = \rho^2 + F\left(\rho\right)^2 \;, \label{fpd}
\ee
where $F\left(\rho\right)^2 =  \frac{4\kappa}{3}  g^{ij} D_i Z \overline{D}_j \overline{Z}$. 

We can now utilise (\ref{fpd}) in (\ref{defphprsim}) to obtain an upper bound on the proper field distance $\Delta \phi$. Since we are interested in an upper bound, we are free to simply drop the $F^2$ term in (\ref{fpd}). However, this term has a crucial role as a regulator for the $\rho \rightarrow 0$ regime. Since $h_i g^{ij} h_j$ is a norm for a vector with a positive definite metric, it can only vanish at singular points in moduli space away from the large volume regime. Therefore (\ref{fpd}) implies that as $\rho \rightarrow 0$, within the large volume geometric regime, $F$ must tend to a minimum finite value. Let us denote the minimum value of $h_i g^{ij} h_j$, for the interval $\rho_i \leq \rho \leq \rho_f$,  by $\rho_M^2$. Let us also assume, for simplicity and with generality, that $\rho$ is positive and that $\rho_i \leq \rho_M \leq \rho_f$. Then we can write 
\be
\Delta \phi \leq \int_{\rho_i}^{\rho_M} \frac{1}{\rho_M} d\rho + \int_{\rho_M}^{\rho_f} \frac{1}{\rho} d\rho = 1 - \frac{\rho_i}{\rho_M} + \ln \left(\frac{\rho_f}{\rho_M}\right) \;. \label{loggr}
\ee
Therefore we find that the proper field distance grows at best logarithmically with $\rho$ for  $\Delta \phi \geq 1$, with a prefactor which is smaller than, or equal to, one. This holds for any linear combination of moduli for any Calabi-Yau. The result provides further evidence for the conjecture made in \cite{Baume:2016psm} that field distances grow logarithmically for $\Delta \phi \geq 1$. 

Utilising the results in \cite{Palti:2008mg}, it can be checked that the first $\alpha'$ correction to the metric, which for ${\cal N}=2$ appears at $\alpha'^3$, modifies the prepotential in such a way that $-\frac12e^{-K}{\cal I}^{ij} \neq  g^{ij}$.\footnote{This also shows that this equality does not hold for an arbitrary ${\cal N}=2$ system.} Therefore the bound $h_i g^{ij} h_j  \geq  \rho^2$ is modified. It would be interesting to see what results could be obtained away from the large volume regime, though it appears unlikely that this regime can support large field variations. 

\subsubsection{Moduli of Calabi-Yau Orientifolds}
\label{sec:ori}

The results on moduli so far applied to Calabi-Yau compactifications. Another often-studied class of compactifications are Calabi-Yau orientifolds. These compactifications preserve ${\cal N}=1$ supersymmetry but they are not completely general ${\cal N}=1$ theories and still support an underlying ${\cal N}=2$ structure.

We consider orientifold projections in type IIA since they can be written in a more universal manner. However, they capture the physics of their type IIB duals (as well as heterotic duals and possible F-theory uplifts). The effective action and ${\cal N}=2 \rightarrow {\cal N}=1$ projection are described in detail in \cite{Grimm:2005fa}. The action on the Kahler moduli is very simple, it is a truncation of the index range of the fields. Therefore the ${\cal N}=2$ analysis carries through unchanged. The complex-structure moduli sector is more complicated because it is a projection from a special quaternionic manifold spanned by the ${\cal N}=2$ hypermultiplets, which includes a special Kahler submanifold spanned by the complex-structure moduli,  onto a different Kahler sub-manifold spanned by the ${\cal N}=1$ chiral multiplets (which is in fact special Lagrangian). The projection acts by splitting the hypermultiplets into two type of chiral multiplets whose scalar components we denote
\be
N^k = \frac12 \xi^k + i l^k  \;,\;\; T_{\alpha} = i \tilde{\xi}_{\alpha} + \tau_{\alpha} \;.
\ee
The index range of $k$ and $\alpha$ sum to that of the $I$ on the original ${\cal N}=2$ special Kahler manifold. If the $0$ index value of the $I$ lies in the $k$ range then these are dual to type IIB O3/O7 orientifolds, while if the $0$ index is in the $\alpha$ range they are dual to type IIB O5/O9 orientifolds. The fields $l^k$ and $\tau_{\alpha}$ are projections from the ${\cal N}=2$ special Kahler periods such that
\be
l^k = \mathrm{Re} \left(C X^k\right) \;,\;\; \tau_{\alpha} = -2 \mathrm{Re} \left(C F_{\alpha} \right)\;. 
\ee
Here $C$ is the compensator field which is related to the four-dimensional dilaton $D$ through $C = e^{-i\theta - D+ \half K}$, where $K$ is the appropriately truncated complex-structure Kahler potential (\ref{N2KahlerPot}) and $\theta$ is an angle defining the calibration of the special Lagrangian sub-manifold in the quaternionic space. 

The ${\cal N}=1$ theory inherits a truncated ${\cal N}=2$ structure. The truncation is imposed through 
\be
\mathrm{Im} \left( C X^k \right) = \mathrm{Re} \left( C F_k \right) =  \mathrm{Re} \left( C X^{\alpha} \right) =  \mathrm{Im} \left( C F_{\alpha} \right) = 0 \;.
\ee
We can therefore write down a projection of equation (\ref{si}). We consider a linear combination of the moduli fields
\be
\rho = q_k l^k + \frac12 p^{\alpha} \tau_{\alpha} \;.
\ee
The projection of (\ref{si}) then gives, after some calculation (utilising primarily appendices B and C of\cite{Grimm:2005fa}), 
\be
\left| \left( q_k, p^{\alpha} \right) \right|^2 = \rho^2 + G\left(\rho \right)^2 \;.
\ee
Here $\left| \left( q_k, p^{\alpha} \right) \right|^2$ denotes the norm of the linear combination vector as appearing in (\ref{defphprsim}), and $G\left(\rho \right)^2$ denotes a positive definite contribution. We therefore obtain again a logarithmic growth bound (\ref{loggr}).

\subsection{Masses of Towers of States}
\label{sec:masssta}

Having established the logarithmic behaviour of the field distance in the moduli, we can turn to the dependence of towers of states on the field distance and therefore to the RSC (\ref{RSC}). This dependence will also determine the exponent factor $\alpha$ in the RSC (\ref{RSC}).\footnote{It is worth noting that the result (\ref{loggr}) does not imply that $\alpha \geq 1$ in the RSC since the mass dependence on $\rho$ is an additional ambiguity.} 

While we established that the dependence of $\Delta \phi$ on $\rho$ is logarithmic for any values of the the $h_i$, to prove the Refined Swampland Conjecture we need to restrict to the case $h_i \geq 0$. This does not mean that for some negative $h_i$ the RSC can be avoided, only that the form of the inequality (\ref{loggr}) only allows for a strict proof for positive $h_i$. With this restriction, let us denote the index choice $i=M$ as the one for which the term $h_Mt^M$ is the largest one in $\rho_f$. Then we can write
\be
\rho_f \leq n h_M t_{(f)}^M \;, \;\; \rho_i \geq h_M t_{(i)}^M  \;, 
\ee
where $t^M_{(i)}$ denotes the initial value of $t^M$,  $t^M_{(f)}$ its final value, and $n$ is the number of fields appearing in $\rho$. Therefore we have
\be
\frac{t^M_{(f)}}{t^M_{(i)}} \geq \frac{1}{n} \frac{\rho_f}{\rho_i} \geq \frac{1}{n} e^{\Delta \phi'} \;,
\ee
where we define $\Delta \phi' = \Delta \phi - 1 + \frac{\rho_i}{\rho_c}$.
We have therefore established that at least one of the $t^i$ increases exponentially in $\Delta \phi'$. 

Within a string theory compactification setting the fields $t^i$ control the sizes of cycles. There is an infinite tower of modes whose mass decreases as a power of the size of these cycles. For example, in type IIA string theory if the $t^i$ are Kahler moduli they directly measure the volumes of cycles and so the mass of KK modes. If they are complex-structure moduli then they are mirror dual to Kahler moduli in type IIB string theory which measure the size of cycles and therefore have KK towers associated to them. The precise dependence of the KK mass on $t^i$ depends on the extra dimensional geometry as well as which geometric quantity the $t^i$ are measuring. For a torodial setup we have $M^i_{KK} \sim \frac{1}{t^i}$. But more generally we can parameterise it as
\be
M^i_{KK} \sim  \left(t^i\right)^{-\alpha} \;,
\ee
so that
\be
\frac{M^i_{KK,(f)}}{M^i_{KK,(i)}} \leq n^{\alpha}e^{-\alpha\Delta \phi'} \;. \label{mkkmav}
\ee
The exponential dependence on $\Delta \phi$ of an infinite tower of states matches the RSC. 

The KK masses $M^i_{KK}$ dependence on the $t^i$ is expected to take the rough form of $\kappa^{-\frac12}{\left(t^i\right)^{-\frac{1}{d}}}$ (see for example \cite{Conlon:2005ki} for an analysis). The first factor comes from the modification of the string scale relative to the Planck scale. The second factor is the inverse length scale of the cycle, where $d$ is the dimension of the cycle.\footnote{Note that the $N^k$ moduli in section \ref{sec:ori} come from the reduction of the NS form $B_2$ rather than $J$. However, they are still expected to control the mass of an infinite tower of states in a string theory setup since they contribute to the `stringy' volume of the cycle.} So, say for IIA, we therefore expect a rough bound of $\alpha \geq \frac12$ for electric (two-cycle) moduli and $\alpha \geq \frac14$ for magnetic (four-cycle) moduli. However, it is difficult to prove a general statement on a lower bound for $\alpha$ for general Calabi-Yau and Calabi-Yau orientifold compactifications. 

There is another infinite tower of states which are the wrapped branes on the cycles.\footnote{Note that there is a subtlety in whether to count the multiply wrapped branes as different states or not. This is a question of the Calabi-Yau geometry to do with whether there is an appropriate representative cycle in the homology class of each wrapping. We assume that this holds and that there is indeed such a tower of states. This statement is likely to depend on the region in field space we are in. Looking at two extreme limits: in the case of a conifold, $t^i \rightarrow 0$, it is expected that only one state becomes massless \cite{Strominger:1995cz}. While in the decompactification limit, which is the one of relevance here, it is more natural to expect an infinite tower of states.} These are particles for even cycles in IIA and odd cycles in IIB. For odd cycles in IIA the relevant states are strings whose tension is controlled by the cycle, but we will henceforth consider the particles case. When there is a large number of them they can be described as a classical charged extremal black hole with a mass given by (\ref{admmass}). The mass of a small number of probe wrapped branes is still given by the central charge $\left|Z\right|$ with small charge vectors. Looking at unit charged vectors, and applying the cubic prepotential as in (\ref{cupre}), we have electric and magnetic mass scales
\be
m_e^0 = \sqrt{\frac{3}{4\kappa}} \;,\;\; m_e^i = \sqrt{\frac{3}{4\kappa}} t^i \;,\;\; m^m_i = \sqrt{\frac{3}{4\kappa}} \frac{\kappa_i}{2} \;,\;\; m^m_0 = \frac{1}{6}\sqrt{\frac{3\kappa}{4}} \;. \label{embrato}
\ee
Since $t^M$ must appear at least linearly in $\kappa$ we have
\be
m_0^e \leq  \sqrt{\frac{3}{4t^M_{(f)}}} \leq  \sqrt{\frac{3nh_M}{4\rho_f}} \leq \sqrt{\frac{3n}{4}} e^{-\frac{\Delta \phi'}{2}} \;. \label{bhbma}
\ee
In the last inequality we used the fact that $h_M\leq 1$ and that for $h_i \geq 0$ we have that $\rho_i \geq 1$ in the large volume regime. Note that also, either the mass scale $m_e^i$ or $m^m_i$, depending on how $t^M$ appears in $\kappa$, have exponentially decreasing bounds.  

The towers of states (\ref{embrato}) are particularly interesting because they are in some sense more general than the KK states. This is because they can be argued to be present generally from an ${\cal N}=2$ supergravity perspective without referring to a string compactification. Indeed, they can be associated to the towers of states of the Lattice WGC \cite{Heidenreich:2015nta,Heidenreich:2016aqi}. They are therefore important with respect to the generality of our results. Within a string compactification setting, however, they are, at least in a generic large volume limit, heavier than the KK modes and so we expect that practically the strongest bound on the effective field theory comes from KK towers. 

Note that (\ref{bhbma}) is a bound on the mass rather than on the mass variation as in (\ref{mkkmav}). The reason for this is that a variation of $t^{M}$ may not influence $\kappa$ very much if the other moduli were much larger than $t^{M}_{(f)}$. Therefore, we can not prove that the states become exponentially lighter, but only that their mass is restricted by an exponentially decreasing bound. This is a slightly weaker statement than the RSC. It is not clear which version should be imposed generally: that the mass scale decreases exponentially or that the bound on the mass scale decreases exponentially. Of course, for all practical purposes, the difference is not of great importance in that if the tower mass scale was very low to start off with it would only place a stronger restriction than the RSC on the breakdown of the effective field theory.\footnote{It is analogous to the statement that forbidding a monopole from being a black hole, as in \cite{ArkaniHamed:2006dz}, could be resolved by some other states, not necessarily gravitational, before the collapse to a black hole. As is the case for an $SU(2)$ monopole.} But it is worth pointing out this subtlety.

\section{Gravity as the Weakest Force}
\label{sec:refgrw}

In section \ref{sec:n2ebh} we reached the conclusion that the gauge forces should act more strongly than the gravitational and scalar forces combined. In this section we consider the relative magnitude of the scalar and gravitational forces. We follow similar methodology to the gauge force analysis where we first establish relations between the forces acting on the WGC states within the ${\cal N}=2$ context. We will then propose general relations based on capturing the relevant physics, however, this generalisation is much more complicated than in section \ref{sec:n2ebh} and it is less clear if it holds. In section \ref{sec:distmod} we established connections between the gauge force statement and distances in moduli space and the RSC. It is therefore natural to explore if a statement on scalar forces also has implications for the field space. We show that indeed ties to the RSC can be established. 

\subsection{A Scalar Weak Gravity Conjecture}

In the ${\cal N}=2$ supergravity setting the key equation for the gauge forces was (\ref{si}). This was interpreted as the self interaction of the WGC states. We now would like to consider interactions between different states. The relevant relation generalising (\ref{si}) reads (see \cite{Andrianopoli:2006ub,Ritz:2001jk} for useful texts)
\be
\left|Z\right|\left|Z'\right| + \mathrm{Re}\left(4g^{ij} \partial_i \left| Z' \right| \overline{\partial}_j \left| Z \right| \right) = {\cal Q}{\cal Q}' \mathrm{Re}\left(\frac{Z \overline{Z}'}{\left|Z \overline{Z}' \right|} \right) - \frac12 \left(q_I  p'^{I} - q'_I p^I \right) \mathrm{Im}\left(\frac{Z \overline{Z}'}{\left|Z \overline{Z}' \right|} \right)\;.\label{noforce}
\ee
The first term in (\ref{noforce}) is the gravitational force and the second is the force mediated by the scalars. The gauge force is given by ${\cal Q}{\cal Q}'$. The last term is only non-vanishing if the interacting states are not mutually local. We henceforth restrict to mutually local states and so take it to vanish. Now the important point is that we can consider two states with a vanishing gauge force between them ${\cal Q}{\cal Q}' =0$. In this case we see that the scalar forces cancels the gravitational force. Therefore for states with vanishing vector interactions the scalar forces act repulsively. 

In the case of gauge forces we were able to argue that the ${\cal N}=2$ results should hold generally due to the interpretation as forbidding gravitationally bound states. We can attempt to apply the same logic also for the relative magnitude of the scalar and gravitational forces in the case when the gauge vector forces vanish. However, we will see that there is an important difference in this case. The absence of a stable gravitationally bound state requires that the scalar forces act stronger than gravity. We can therefore conjecture that if we consider two WGC states of mass $m$ and $m'$, which are mutually local and have vanishing gauge interactions, then, for the general theory (\ref{genaction}), the scalar forces must act repulsively and at least as strong as gravity
\be
-g^{ij} \left(\partial_{t^i} m\right)\left( \partial_{t^j} m' \right) \geq m m' \;. \label{2sswgc}
\ee
It can also be seen from (\ref{noforce}) that if the gauge force is non-vanishing, then its magnitude is equal to the scalar and gravitational forces combined, which can be taken as a conjecture generalising (\ref{gnsWGC}). Note that we observe already a striking statement: the existence of two $U(1)$ gauge fields requires the existence of a scalar field. The role of the field is to stop the gravitational bound state made from the possible orthogonal charge choice.

Note that we consider mutually local states because the ${\cal N}=2$ formalism only makes sense in such cases. However, if we consider a purely magnetic state and a purely electric one, then they will again not exert any gauge force. This is clear even if we can not describe them simultaneously as local states in a field theory.\footnote{Note however that the structure of the gravitationally bound state is more complicated than that of two electric states which are orthogonal with multiple $U(1)$s because a gravitational orbit will induce a relative velocity of the particle and therefore a $U(1)$ force.} Applying the same logic therefore implies that again the scalar force must act repulsively and stronger than gravity.  This also implies that the presence of a scalar field is required even by a single gauge field to stop stable gravitationally bound dyonic states.  

Having stated the natural generalisation of (\ref{noforce}) as (\ref{2sswgc}), the arguments for this generalisation are far less clear than in the gauge case in section \ref{sec:n2ebh}. The first fundamental difference is that while the analysis in section \ref{sec:n2ebh} was regarding the implications of the presence of massless scalar fields, here we are requiring the presence of a scalar field, which is a much stronger statement. This also means that we must state how a mass for this field can affect the setup. The idea is that the scalar field can be massive. Then (\ref{2sswgc}) ensures that if a bound state exists then its radius is set by the inverse mass of the scalar field. This is analogous to the discussion in section \ref{sec:n2ebh} regarding the meaning of the WGC when the gauge field is massive. 

The second fundamental difference is in considering a bound state of two orthogonal states compared to a bound state of the same particle. The difference is in how a tower of states could arise. Consider, for illustration purposes, the case of a single electrically charged state, denoted $\left(1,0\right)$, and a single magnetically charged state, denoted $\left(0,1\right)$. The same analysis applies to two orthogonal electric states in the case of multiple $U(1)$s. Then let us violate (\ref{2sswgc}) and see if a tower of stable gravitationally bound state can be constructed.  

One way to create a tower is to consider forming a bound state from the single charged states, so of charge $\left(1,1\right)$. And then considering an orthogonal dyonic state to it. Or in other words, performing an electric-magnetic rotation so the bound state is purely electric and then adding a magnetic state in this new frame to it. However, this is unsatisfactory for a number of reasons. One reason is that it is not clear that with an arbitrary gauge kinetic matrix this can be done since the charges are quantised while the moduli are continuous. Another reason is that the next state in the tower has twice the charge of another state in the theory and so can decay to two such states. This must somehow be forbidden for the tower to be stable, and it is unclear to us how. Yet another reason is that the original bound will have a dipole $U(1)$ moment due to its finite size, which would induce a gauge interaction. 

Another way to consider forming a tower of stable states is to consider a tower of $\left(1,1\right)$ bound states. These states will feel a gauge repulsion, but this could be overcome by the attraction. If the $\left(1,0\right)$ and $\left(0,1\right)$ states are precisely extremal, as in the ${\cal N}=2$ setting, then a violation of (\ref{2sswgc}) would imply a tower of stable states. More generally we should require a constraint as in (\ref{gnsWGC}) for the $\left(1,1\right)$ bound state itself
\be
{\cal Q}_B^2 \geq m_B^2 + g^{ij} \mu_i^B \mu_j^B \;,\label{2sswgc2}
\ee
where the $B$ subscript denotes that this is a bound state. If (\ref{2sswgc2}) holds then even if (\ref{2sswgc}) is violated there would still not be a tower of stable gravitationally bound states. 

In the absence of scalar fields, if the constituents of the bound state are precisely extremal, then (\ref{2sswgc2}) is violated and we have a stable tower. Heavy states in this tower are extremal black holes along this charge direction. Requiring them to decay, or that the tower of stable states is absent, means that the constituents can not be precisely extremal but must be slightly super-extremal. This is the convex hull condition as studied in \cite{Cheung:2014vva, Rudelius:2015xta,Brown:2015iha}. 

If (\ref{2sswgc2}) is violated and we have a tower, and also stable black holes, then another solution is to propose that this must decay and therefore there should exist a super-extremal particle for each charge lattice, this is the reasoning for the Lattice WGC \cite{Heidenreich:2015nta}. We see that the scalar fields offer an alternative, we can demand that (\ref{2sswgc}) or (\ref{2sswgc2}) must hold once all the scalar fields are accounted for, including massive ones. Then the bound states are such that their radius is limited by the mass of the scalar. This may be less problematic from a quantum gravity perspective.

Note that if we retain the Lattice WGC, and consider the lattice of charges to be populated by states, then we must consider forming a tower from bound states of $\left(m,0\right)$ with $\left(0,n\right)$. By themselves each of $\left(m,0\right)$ and $\left(0,n\right)$ are not expected to be stable. If the charge to mass ratio in the tower of states of the Lattice WGC decreases as we move up the ladder then we have a decay channel $\left(0,n\right) \rightarrow n\left(0,1\right)$. However, a gravitationally bound state can be stable against this decay. This depends on whether the increase in the charge to mass ratio due to the gravitational binding energy is larger than the decrease due to having a higher state in the tower. Understanding this would require understanding the structure of the tower and of the bound state better. 

In summary, it is not clear what is a possible generalisation of (\ref{noforce}), and what are the resulting implications. We considered (\ref{2sswgc}) and (\ref{2sswgc2}) as possibilities. However, the uncertainty in the stability of the tower of bound states, and in what problems such a tower causes, means that their generality is on less firm footing than that of (\ref{gnsWGC}).

We are also interested in the relative magnitude of the scalar and gravitational forces for the interaction of a state with itself. The relevant ${\cal N}=2$ equation here is \cite{Ceresole:1995ca}
\be
{\cal Q}^2\left(F\right) = \left|Z\right|^2 - g^{ij} D_i Z \overline{D}_j \overline{Z} \;. \label{si2}
\ee
Here ${\cal Q}^2\left(F\right) $ is defined in the same way as in (\ref{qdefq}) but with ${\cal N}_{IJ} \rightarrow F_{IJ}$. In the same way that we interpreted (\ref{si}) as a bound on the sum of the scalar and gravitational forces, (\ref{si2}) gives information on their relative magnitude. The matrix $I_{IJ}$ is negative definite. The matrix $\mathrm{Im}\left(F\right)_{IJ}$ has $n_V$ strictly positive eigenvalues and one strictly negative eigenvalue. This means that if we consider the WGC states, there is a basis where $n_V$ of them have scalar forces acting strictly stronger than gravity, and one of them has gravity acting strictly stronger. The odd one out is due to the graviphoton which has no scalar superpartners. In other words, we can say that for each scalar field there is one WGC state for which gravity is the weakest force.\footnote{There are $2n_V$ real scalar fields, and $n_V$ electric plus $n_V$ magnetic WGC states.}  

We can formulate (\ref{si2}) generally as 
\be
g^{ij} \left(\partial_{t^i} m\right)\left( \partial_{t^j} m \right) > m^2 \;, \label{wgcs}
\ee
for the general theory (\ref{genaction}). In the ${\cal N}=2$ case the spectrum of states satisfying this was such that there was one (electric and one magnetic) state which violated (\ref{wgcs}) and all the others satisfied it. It appears that this structure is general if the states are precisely extermal. Consider only electric states for now, and for simplicity just one canonically normalised scalar field so that  (\ref{2sswgc}) reads
\be
 \left|\partial_{t} m\right|\left|\partial_{t} m' \right| \geq m m' \;. \label{2sswgcsim}
\ee
This must hold for all state with vanishing gauge interactions, which are therefore orthogonal with respect to the matrix ${\cal M}$ in (\ref{Mdef}). Now say that one state was such that $\left|\partial_{t} m\right| < m$, then all the other orthogonal states must have $\left|\partial_{t} m\right| > m$. Therefore we deduce that at most one state can violate a non-strict inequality version of (\ref{wgcs}), and that if it does then (\ref{wgcs}) becomes a strict inequality for all the other states. 

The generality of (\ref{wgcs}), away from the ${\cal N}=2$ framework, cannot be directly deduced by thinking about the existence of bound states since both the scalars and gravity act attractively. It is possible that it can be deduced from  (\ref{2sswgc}) or (\ref{2sswgc2}), but this is not clear. We can propose that (\ref{wgcs}) holds generally, but leave building more evidence for this for future work.

We can denote the statements (\ref{2sswgc}), or more generally (\ref{2sswgc2}),  and (\ref{wgcs}) as the Scalar WGC. We repeat, that the argument for them is much less strong than that of (\ref{gnsWGC}). With respect to the generality of the conjecture, there are two natural possibilities. The first is that it should hold as a statement about the scalar interaction of the WGC states associated to gauge fields. We can term this the Gauge-Scalar WGC. The second is that it should hold completely generally even in the absence of gauge fields, we can term this the General Scalar WGC. This is the more general statement that gravity truly is the weakest force, so for each scalar force there is a state on which gravity acts more weakly. We can again motivate it in terms of forbidding gravitationally bound states, however, in the absence of an associated gauge symmetry it is unclear what could lead to the stability of such states. It is possible that one could associate some, at best approximately conserved, charge to the scalar fields. For example, a shift symmetry. However, in the absence of a solid argument for the stability of bound states coupled to scalar fields only, the evidence for the General Scalar WGC remains rather weak. Due to the combination of the Scalar and Gauge WGCs being the single statement that gravity is the weakest force we will often refer to them both jointly as the WGC. The rest of the analysis in this section holds for either version of the scalar conjectures. 
 
Before proceeding it is worth mentioning another interesting identity in ${\cal N}=2$
\be
g^{ij}D_{i} \overline{D}_{j} \left| Z \right|^2 = n_V \left| Z \right|^2 +  g^{ij}D_i Z\overline{D_i Z} \;. \label{wgcquZ}
\ee 
We can interpret this as a relation between the four-point coupling, the mass and the three-point coupling of the WGC states to the scalar fields. It is possible to phrase it as a bound that the four-point coupling for two scalar fields to two WGC states should be larger than the coupling of two gravitons to two WGC states.\footnote{The factor $n_V$ appears because the WGC states only couple to one combination of gauge fields but to all the moduli. This can be seen by restricting to electric charges so that the WGC state couples to just one linear combination of axions, and calculating the relation for the axions $b^i$ and moduli $t^i$ separately. The factor of $n_V$ the only appears in the coupling to the moduli $t^i$.} It would be interesting to see if there is a general interpretation for the physics of equation (\ref{wgcquZ}).\footnote{We can formulate a Scalar WGC based on (\ref{wgcquZ}) which would suggest $n m^2 + g^{ij}\left(\partial_{t^i} m\right)\left( \partial_{t^j} m \right) \leq \frac12 g^{ij} \nabla_{t^i} \nabla_{t^j} m^2$, where $n$ is the number of fields coupling to the WGC state.} 

\subsection{The Refined Swampland Conjecture and Gravity as the Weakest Force}

The bound (\ref{wgcs}), which states that gravity is the weakest force, can be thought of as differential equation in the mass of the WGC states. Regardless, of whether we are at this point able to strongly motivate (\ref{wgcs}), it is informative to consider its implications. In general, these are complicated coupled non-linear differential equations. But to illustrate the key point consider a simple theory of a single canonically normalised scalar field $t$. Then, for all but one of the states we have,
\be
\left|\partial_t m\right| > m  \;. \label{dmt}
\ee
Consider a power-law form $m=t^p$, then (\ref{dmt}) gives
\be
\left|p\right|  > t \;.
\ee
This will be violated for large enough $t$. Indeed, to satisfy the inequality for arbitrarily large $t$ the mass must be an exponential 
\be
m = e^{-\alpha t} \;, \label{rscgw}
\ee
with $\left|\alpha\right| > 1$. Therefore for large $t$ the behaviour of the mass of the WGC states must be exponential, but this is precisely the Swampland Conjecture (if we also ask that the lattice of charges is populated) \cite{Ooguri:2006in}. Indeed, we see that the behaviour asymptotes to exponential quickly for $t > 1$, so that we recover the Refined Swampland Conjecture.\footnote{The behaviour where the power in a power law must increase with the field displacement is the same as that found in \cite{Klaewer:2016kiy}.} We also find a lower bound on $\alpha$, which has important phenomenological implications. Note that (\ref{2sswgc}) implies that the sign of $\alpha$ is opposite for states with vanishing gauge interactions. This means that there is always one state for which $\alpha$ is positive and so its mass decreases. 

In other words, we find some evidence towards the idea that the RSC can be understood as the statement that there must exist a state on which the force that the scalar field mediates acts more strongly than gravity. So the RSC and WGC are not just related by supersymmetry, but they are together forming the general statement that gravity is the weakest force. 

It is interesting to consider how this conclusion is compatible with periodic axions. We can consider the ${\cal N}=2$ framework and take the $b^i$ to be the axions. Then the exponential behaviour (\ref{rscgw}) is incompatible with the axion periodicity. How this is resolved can be understood as follows. Consider applying (\ref{dmt}) to a state of charges $\left(q,p\right)$. Now if we take $b^i = n + b^{i'}$, with $b^{i'} \leq 1$ and $n$ being some integer, then we can instead consider a different state with charges $\left(q',p'\right)$ such that $m\left(b^i,p,q\right) = m\left(b^{i'},p',q'\right) $. The conjecture is that there must exist a state for which (\ref{dmt}) holds, but this does not have to be the same state for all values of the field.\footnote{There is a subtle point here because the periodicity of the $b^i$ requires the full charge lattice $\left(q_I,p^I\right)$ which includes the graviphoton charge. This means that the re-arrangement of states is such that generically the states for which gravity acts weaker than the scalars, and the one state on which this is not true, get mixed up. Nonetheless, there is always a basis at each point in field space for which $2n_V$ states have gravity as the weakest force.} This means that effectively, by choosing different states as we move around the axion field space, we can consider $b^i \leq 1$, and so there is no required exponential behaviour (\ref{rscgw}). Note that this argument ties the existence of periodic fields, axions, with the WGC states populating a full lattice. This can be viewed as evidence for a scalar version of the Lattice WGC.\footnote{It also suggests that the Lattice WGC for gauge fields \cite{Heidenreich:2015nta,Heidenreich:2016aqi} may be understood in terms of the periodic structure of expectation values of line operators.} 

We showed how axions escape the exponential behaviour through an infinite tower of states. For a monotonic function $m\left(t\right)$ we require such an infinite tower. If we allow for an oscillatory function it is possible to satisfy (\ref{dmt}) with only two states interchanging their role as the WGC state.\footnote{I thank Arthur Hebecker for pointing this out. }  Note that in string theory this is the relevant case for axion fields which only appear through world-sheet instanton corrections (so exponentially inside the periods). However, the period of the oscillations must be less than one. The period could be made longer by including more and more states, until we reach the infinite tower of the non-oscillatory case. It is interesting to see a connection between many states and field distances. Note also that the field distance excursion need not be tied to the period length in an axion monodromy type scenario, and so such scenarios are not constrained by this. 

Therefore, we can not say for certain, even for the simple one-field case, that the mass must be exponential. We can make the statement that if $\frac{\left|\partial_t m\right| }{m}$ increases monotonically (or stays constant) then it must be an exponential with exponent $\left|\alpha\right| > 1$. We can also expect exponential behaviour if the mass of the tower of states behaves as $m_n\left(t\right) = n f\left(t\right)$, where $f$ is an arbitrary function and $n$ is an integer denoting the state in the tower. Then since $\frac{\left|\partial_t m_n\right| }{m_n}$ is independent of $n$ it is not possible for the states to replace each other as the WGC states. A thorough proof or analysis of when the exponential behaviour is present requires more work but, having stated the caveats, the exponential behaviour (\ref{rscgw}) is certainly compelling in terms of evidence towards a connection between the Scalar WGCs and the RSC.

\section{Summary}
\label{sec:summary}

We studied aspects of the Weak Gravity Conjecture in the presence of scalar fields. We utilised the structure of ${\cal N}=2$ black holes to formulate the WGC generally for arbitrary scalar field-space metric and gauge kinetic function. The conjecture can be phrased as the statement that there must exist a particle on which the gauge force should act stronger than gravity and the scalar forces combined. The underlying principle is that the WGC particles should not form a tower of stable gravitationally bound states. There are arguments for why this should be the case in \cite{ArkaniHamed:2006dz,Cottrell:2016bty}. We did not develop these further but took the absence of such states as an assumption and used it to formulate the conjecture. The results are therefore reliant on this assumption and it would be very interesting to develop the quantum gravity reasoning for the absence of such towers more rigorously. The conjecture would also follow from requiring the decay of extremal black holes if they exhibit certain structures which are present in an ${\cal N}=2$ supersymmetric context. There are examples and evidence towards this possibility within the so-called `fake superpotential' formalism \cite{Denef:2000nb,Ceresole:2007wx}. The conjecture exhibits the interesting property that it remains non-trivial even for states with a very light mass, and therefore has an interesting infrared limit.

The WGC bound is marginally satisfied by BPS states in ${\cal N}=2$ supergravity. This can be shown utilising an ${\cal N}=2$ identity. We showed that this identity can be utilised to extract information on the behaviour of field distances in scalar field spaces. As an application we presented a proof that for any linear combination of moduli in Calabi-Yau or Calabi-Yau orientifold compactifications, the proper field distance grows at best logarithmically with the moduli for super-Planckian distances. We also identified infinite towers of states whose mass decreases exponentially as a result. This general proof presents new evidence for the Refined Swampland Conjecture developed in \cite{Baume:2016psm,Klaewer:2016kiy}. 

We showed in the ${\cal N}=2$ supergravity setting that scalar field forces also act at least as strongly as gravity on the WGC states associated to the gauge fields. We considered two statements generalising this, (\ref{2sswgc}) and (\ref{wgcs}). The two statements together were termed the Gauge-Scalar WGC. However, the evidence for this in terms of a tower of stable gravitationally bound states is weaker than in the case where we consider only a single state forming a tower. Establishing the existence of a tower, or other evidence for the Gauge-Scalar WGC, away from the ${\cal N}=2$ setting, requires further work. 

We also formulated a General Scalar WGC which states that the property of the WGC states, of having scalar fields act stronger on them than gravity, is independent of their connection to the gauge fields. So that for every scalar field there must exists a state on which gravity acts weaker than the scalar field. This amounts to directly imposing that `gravity is the weakest force'. It can be motivated by forbidding gravitationally bound states, but in the absence of the gauge symmetry it is even more unclear how their stability is ensured and therefore is less motivated than the Gauge-Scalar WGC.

We showed that the Scalar WGCs naturally lead to the behaviour of the mass of the WGC states to be exponential in the scalar field expectation value, as in the RSC. This introduces a candidate general physical principle behind the RSC. We were unable to show that the RSC is implied by the Scalar WGCs with generality due to the latter being formulated as complicated coupled non-linear differential equations for the mass. Even in the simple case of a single scalar field we showed that there are ways to avoid the exponential behaviour. One way is if there are a large number of states which play the role of the WGC state at different points in field space. Applying this to axions we argued that in fact there must be an infinite tower of such states. Another possibility is if the WGC mass is oscillatory. 

In the single scalar field case we were able to show, up to certain assumptions, that the exponential behaviour is such that the exponent is bound to be larger than one $\left|\alpha\right| > 1$. This bound can have important implications for large field inflation because it implies a lower bound how fast the tower of WGC states becomes light for super-Planckian distances.\footnote{See also \cite{Blumenhagen:2017cxt} for a quantitative study of a bound due to such physics.} This in turn limits the energy scale of the effective theory of inflation and therefore places a direct bound on the magnitude of the primordial gravitational waves that could be produced. The most conservative application of this bound does not yet yield numbers comparable with current experiments. For example, if we take the initial mass scale of the tower of states to start at the Planck scale, and impose that the Hubble scale during inflation should be lower than the tower mass scale, then we find for the tensor-to-scalar ratio $r<1$. This conservative bound can be easily sharpened by additional restrictions on the effective theory. Of course, this assumes the possible application of the exponential behaviour to the inflaton, which is subject to the assumptions and subtleties discussed in this work.

If true, the Gauge-Scalar WGC has some striking conclusions. It implies that the existence of gauge fields requires the existence of scalar fields. Otherwise, there would be nothing to stop forming gravitationally bound states when the gauge interaction vanishes. Further, the couplings of the gauge fields are tied to the cubic coupling of the scalar fields, through the properties of the WGC states. We are seeing the emergence of supersymmetric vector multiplets. This leads to the striking possibility that the WGC, or similar general reasoning about quantum gravity, could imply the existence of high scale supersymmetry.

In M-theory there are no constant coupling parameters, so that all the parameters in four-dimensions are functions of scalar fields. This means that, in contrast to gauge fields, every state in the theory must couple to some scalar field. It therefore allows for the possibility of a Super WGC, which is that gravity is the weakest force acting on any state. At least this could hold at sufficiently short distance scales above the mass of the scalars. Of course, there is no sufficiently strong evidence for such a strong statement, but we simply want to point out that the existence of scalar field forces means that this possibility is at least not ruled out.

{\bf Acknowledgements:} I would like to thank Arthur Hebecker and Pablo Soler for extremly helpful discussions. I am supported by the Heidelberg Graduate School of Fundamental Physics. 

\appendix

\section{Derivation of Field Space Distance Formula}
\label{sec:derv}

In this appendix we derive the formula (\ref{defphprsim}), which appeared already in \cite{Baume:2016psm}, for the distance in field space along the direction $\rho = h_i t^i$. We consider the kinetic term
\be
{\cal L}_{\mathrm{Kin}} = - g_{ij} \partial t^i \partial t^j \;.
\ee
We now perform a coordinate change to $\sigma_i$ defined by $\partial t^i = M^{ij}\partial \sigma_j$ for some matrix $M_{ij}$ with inverse $M^{ij}$. Note that this is not equivalent to the coordinate change $ t^i = M^{ij}\sigma_j$ since $M_{ij}$ is not constant in general. The kinetic terms now read
\be
{\cal L}_{\mathrm{Kin}} = - g_{ij} M^{ik} M^{jl} \partial \sigma_k \partial \sigma_l \;.
\ee
We now split the $\partial \sigma_i$ into $\partial \sigma_0 = \partial \rho$ and $\partial \sigma_{\lambda}$. We parameterise the matrix $M$ as
\be
M^{i0} = g^{ij}l_{j} \;.
\ee
The $l_j$ are chosen such that 
\be
g_{ij}M^{i0}M^{j\lambda} = g_{ij} g^{il} l_l M^{j\lambda}  = l_j M^{j\lambda} =0 \;. \label{cons1}
\ee
This means that there is no kinetic mixing between $\rho$ and the $\sigma_{\lambda}$. The equations (\ref{cons1}) fix the $l_i$ to be proportional to $M_{0i}$. With the appropriate normalisation we then find
\be
M^{i0} = \frac{g^{ij}M_{0j}}{g^{kl}M_{0k}M_{0l}} \;.
\ee
The kinetic terms then take the form
\be
{\cal L}_{\mathrm{Kin}} = - \frac{1}{g^{kl}M_{0k}M_{0l}} \left(\partial \rho \right)^2  + ... \;,
\ee
where the $...$ denote kinetic terms where $\rho$ does not appear. Now from the definition we have
\be
\partial \rho = M_{0i} \partial t^i \;,
\ee
which implies that $M_{0i}=h_i$, leading to (\ref{defphprsim}).



\end{document}